\documentclass{article}

\PassOptionsToPackage{numbers}{natbib}
 \usepackage[preprint]{neurips_2026}
\usepackage[utf8]{inputenc} % allow utf-8 input
\usepackage[T1]{fontenc}    % use 8-bit T1 fonts
\usepackage{hyperref}       % hyperlinks
\usepackage{url}            % simple URL typesetting
\usepackage{booktabs}       % professional-quality tables
\usepackage{amsfonts}       % blackboard math symbols
\usepackage{nicefrac}       % compact symbols for 1/2, etc.
\usepackage{microtype}      % microtypography
\usepackage{amsmath, amssymb}
\usepackage{graphicx}
\usepackage[table]{xcolor}
\usepackage{algorithm, algorithmic}
\usepackage{pifont}
\usepackage{titletoc}

\graphicspath{{./images/}}

\newcommand{\unc}[1]{{\scriptsize(#1)}}

\title{Accelerating Redshift-Conditioned Galaxy Image Synthesis with One-step Generative Modeling}

\author{%
  Tianyue Yang\\
  The Center for Computational Science\\
  University College London\\
  \And 
   Sandro Tacchella \\
  Cavendish Laboratory \\ Kavli Institute for Cosmology\\
  University of Cambridge\\
 \\ 
  \And
   Xiao Xue\thanks{Corresponding Author.} \\
  The Center for Computational Science\\
  University College London\\
  \texttt{x.xue@ucl.ac.uk} \\ 
}

\begin{document}

\maketitle

\begin{abstract}
Understanding galaxy morphology evolution across cosmic time requires models that can generate realistic galaxy populations conditioned on redshift. In this work, we study efficient redshift-conditioned generative modeling for astrophysical image synthesis using diffusion models and pixel-MeanFlow. We first review the connections between score-based diffusion models, Flow Matching, one-step generative models, and modern diffusion samplers. We then evaluate DDPM, DDIM, DEIS-AB2, DPM++2M, and one-step pixel-MeanFlow on the GalaxiesML-64 dataset using morphology-based metrics, including ellipticity, semi-major axis, S\'ersic index, and isophotal area. Our results show a clear accuracy-efficiency trade-off: standard DDPM sampling achieves the best distributional fidelity but requires high computational cost, while second-order samplers substantially improve efficiency over DDIM. Pixel-MeanFlow enables single-step generation and achieves competitive performance on several morphology statistics, though it remains weaker than many-step DDPM for fine-grained structure. Our results demonstrate that one-step generative models can recover key galaxy morphology statistics at orders-of-magnitude lower computational cost, opening a path toward efficient conditional simulators for large cosmological surveys and simulation-based scientific inference.
\end{abstract}

\section{Introduction}
\noindent \paragraph{Motivation.} Understanding the formation and evolution of astrophysical systems has been an important frontier problem in modern astronomy. Galaxies are dynamical systems that evolve over cosmic time. Directly modeling this evolutionary process is challenging, since we cannot continuously observe a single galaxy over the extremely long timescales on which astrophysical systems evolve. As a result, studies of galaxy evolution are often restricted to learning from discrete snapshots of different systems observed at different cosmic time. Unlike large photometric imaging surveys, such as the Sloan Digital Sky Survey (SDSS)~\cite{2012ApJS..203...21A}, the Dark Energy Survey (DES)~\cite{PhysRevD.105.023520}, the Kilo-Degree Survey (KiDS)~\cite{Kuijken_2019}, and the Hyper Suprime-Cam (HSC) Survey~\cite{2019PASJ...71..114A}, which contain millions or even billions of galaxy images, redshift-matched machine-learning datasets such as GalaxiesML~\cite{do2024galaxiesmldatasetgalaxyimages} are often smaller. Although large spectroscopic surveys such as SDSS and DESI provide redshift measurements for many galaxies, assembling ML-ready datasets requires reliable cross-matching between imaging and redshift catalogs, quality control, and sufficient coverage across the relevant redshift and galaxy-property ranges. In this study, we investigate how generative models can be used to synthesize galaxy images conditioned on cosmic redshift. By generating high-fidelity galaxy populations across cosmic time, this approach helps bridge the observational gap imposed by sparse temporal sampling and provides a scalable framework for modeling how galaxy population statistics vary with cosmic time.~\cite{nguyen2024dreamsmadeemulatingsatellite, Lastufka_2025}.

\paragraph{Generative Modeling.} Generative modeling techniques have long been used for image-based synthesis tasks. Prominent paradigms include Variational Autoencoders (VAEs)~\cite{kingma2013auto}, Generative Adversarial Networks (GANs)~\cite{goodfellow2020generative}, and Normalizing Flows (NFs)~\cite{rezende2015variational}. Since the introduction of Denoising Diffusion Probabilistic Models (DDPMs)~\cite{ho2020denoising}, diffusion-based models have demonstrated superior performance on a wide range of generative vision tasks~\cite{dhariwal2021diffusion}, and have been adapted for scientific tasks in various fields, including weather forecasting~\cite{gao2023prediff, kaifeng2023pangu}, fluid mechanics~\cite{yang2026meno, xue2026uni, ruhling2023dyffusion, ruhe2024rolling, cachay2025elucidated}, and material science~\cite{luo2025crystalflow}. Flow Matching~\cite{lipman2022flow, albergo2023stochastic} (FM) has emerged as a promising alternative formulation to diffusion models, enabling more efficient sampling and more concise training objectives. However, these methods are intrinsically multi-step, and generation quality typically degrades significantly when the number of sampling steps is small. Various approaches have been proposed to address this limitation. For standard diffusion models, the number of sampling steps can be reduced by leveraging more efficient samplers, including Denoising Diffusion Implicit Models (DDIMs)~\cite{song2020denoising}, DPM-Solver++~\cite{lu2022dpmsolverfastodesolver, lu2025dpm}, and Diffusion Exponential Integrator Solver (DEIS)~\cite{zhang2022fast}. Another line of work focuses on few-step generative models. Continuous-time approaches, such as Consistency Models (CMs)~\cite{song2023consistency}, simplified Consistency Models (sCMs)~\cite{lu2024simplifying}, and Shortcut Models~\cite{frans2024one}, leverage self-supervised signals from many-step trajectories to reduce the number of required sampling steps~\cite{lin2026designonestepdiffusionshortcutting}. In contrast, discrete-time approaches typically rely on distribution-matching techniques, with Inductive Moment Matching (IMM)~\cite{zhou2025inductive} serving as a representative example. Recent studies suggest that continuous-time models generally achieve better generation quality~\cite{lin2026designonestepdiffusionshortcutting}.

\paragraph{Generative Modeling for Astrophysics.}
Generative modeling has become an increasingly important tool for astrophysical data analysis, where realistic simulations are often expensive and observational data are high-dimensional, noisy, and affected by complex selection effects. Early work adapted GANs, VAEs, and normalizing flows (NFs) to a range of tasks, including survey augmentation~\cite{lanusse2021deep, bretonniere2022euclid}, emulator-based inference~\cite{mustafa2019cosmogan, perraudin2021emulation}, and out-of-distribution or anomaly detection~\cite{storey2021anomaly, portillo2020dimensionality}. These methods have enabled fast generation of mock observations, data-driven priors for inverse problems, and compact representations of complex astrophysical populations.

More recently, diffusion models~\cite{ho2020denoising} have emerged as a powerful class of generative models for astrophysics, demonstrating strong performance in applications such as realistic galaxy image synthesis, morphology modeling, and cosmological field generation~\cite{lizarraga2025understandinggalaxymorphologyevolution, smith2022realistic, mudur2022can}. Their success is largely due to the strong distribution-matching capability of score-based methods, which can capture multimodal structure and rare events while remaining scalable and largely architecture-agnostic~\cite{dhariwal2021diffusion}. This makes diffusion models attractive for a broad set of downstream applications, including conditional image generation, simulation-based inference, survey systematics modeling, image inpainting, deblending, super-resolution, denoising, and uncertainty-aware forward modeling. Flow-matching (FM) methods have also begun to appear as a more efficient alternative for astrophysical generation tasks~\cite{wildberger2023flow, sayed2025flowlensing}. By learning continuous transport dynamics between simple and data distributions, FM-based approaches offer a flexible framework for conditional generation and fast sampling. 

Despite their promise, both DDPMs and standard FM models are intrinsically multi-step samplers. This can lead to substantial computational cost and wall-clock latency when generating large mock catalogs or applying generative models within iterative inference pipelines. Such costs are especially limiting for modern and upcoming large-scale surveys, where generative models may need to be evaluated millions of times for data augmentation, posterior sampling, likelihood-free inference, or real-time quality-control tasks. This motivates the development of few-step and one-step stochastic generative models for astrophysics. By substantially reducing the number of neural function evaluations, these models could make generative modeling more practical for large-scale deployment while preserving enough fidelity for morphology modeling, survey simulation, anomaly discovery, and fast approximate inference.

\paragraph{Our Methods.} In the one-step and few-step generation regime, MeanFlow models~\cite{geng2025mean} (MF) have emerged as a leading framework. Built upon the Flow Matching formulation, these models learn the average velocity over a time interval. In this study, we explore the capability of the latent-free variant of MF, namely the pixel MeanFlow model (p-MF)~\cite{lu2026onesteplatentfreeimagegeneration}, for continuous redshift-conditioned galaxy generation in astrophysics. Our work builds on prior studies of diffusion-based modeling for understanding galaxy morphology evolution~\cite{lizarraga2025understandinggalaxymorphologyevolution}. Our study provides the following contributions to address these limitations and advance the field:
\begin{enumerate}
    \item We adapt pixel-MeanFlow to the continuously conditioned setting and incorporate modern diffusion training techniques, such as classifier-free guidance, to improve generative performance.
    \item We investigate the efficiency--accuracy trade-off of different few-step diffusion solvers and compare them against pixel-MeanFlow in the astrophysical image-generation setting.
    \item We provide a concise review of existing diffusion models, few-step diffusion models, and sampling techniques, with the goal of inspiring new applications of diffusion-based generative modeling in astrophysics.
\end{enumerate}
To the best of the authors' knowledge, this is the first application of one-step diffusion models, together with a detailed investigation of solver efficiency, in the context of astrophysics. Therefore, our study represents a meaningful advancement in efficient generative modeling for astrophysical applications.

\section{Related Work}\label{sec:related_work}

\paragraph{Diffusion and score-based generative models.}
Diffusion models generate data by learning to reverse a gradual noising process. Early score-based models estimate the score function $\nabla_{\mathbf{x}}\log p(\mathbf{x})$, which avoids the intractable partition function in energy-based models~\cite{doi:10.1126/science.1127647, bruck_1990_m7hdv-r5t49}. Score matching~\cite{hyvarinen2005estimation} and denoising score matching~\cite{vincent2011connection} provide practical objectives for learning this field, while Noise-Conditional Score Networks (NCSNs) extend the idea across multiple noise levels and sample with annealed Langevin dynamics~\cite{song2019generative, song2020score}. DDPMs instead formulate generation as a variational denoising process~\cite{ho2020denoising}. Given a noise schedule $\{\beta_i\}_{i=1}^{L}$, the forward process is
\begin{equation}
    q(\mathbf{z}_i\mid\mathbf{x})
    =
    \mathcal{N}\!\left(
    \mathbf{z}_i;
    \sqrt{\bar{\alpha}_i}\mathbf{x},
    (1-\bar{\alpha}_i)\mathbf{I}
    \right),
    \qquad
    \bar{\alpha}_i=\prod_{j=1}^{i}(1-\beta_j).
\end{equation}
The common $\epsilon$-prediction objective is equivalent to learning a scaled score field through Tweedie's formula~\cite{efron2011tweedie}. Continuous-time extensions unify DDPMs and NCSNs as stochastic differential equations (SDEs), with generation performed by either a reverse-time SDE or an associated probability-flow ODE (PF-ODE)~\cite{song2020score}. Elucidated Diffusion Models (EDMs) further improve practical performance through noise preconditioning, loss weighting, and carefully designed sampling schedules~\cite{karras2022elucidating}.

\paragraph{Flow matching and one-step generation.}
Flow Matching (FM) replaces score estimation with direct velocity-field learning~\cite{lipman2022flow}. It is closely related to normalizing flows and Neural ODEs~\cite{rezende2015variational, chen2018neural, jacobsen2018revnet, lee2025latent, arndt2026invertible}. In Conditional Flow Matching (CFM), one defines an explicit path between noise $\mathbf{x}_0\sim p_0$ and data $\mathbf{x}_1\sim p_1$, for example the optimal-transport interpolation
\begin{equation}
    \mathbf{x}_t=(1-t)\mathbf{x}_0+t\mathbf{x}_1,
    \qquad
    \mathbf{u}_t=\mathbf{x}_1-\mathbf{x}_0.
\end{equation}
The model is trained to approximate this velocity and then sampled by solving an ODE from $t=0$ to $t=1$.

Multi-step samplers remain computationally expensive when the denoising network is large, motivating few-step and one-step models. Consistency Models learn functions that remain invariant along PF-ODE trajectories~\cite{song2023consistency}, with simplified variants improving the parameterization~\cite{lu2024simplifying}. Shortcut Diffusion explicitly enforces consistency between one large step and two smaller steps~\cite{frans2024one}. MeanFlow (MF) generalizes this idea by directly modeling the average velocity over an interval~\cite{geng2025mean}:
\begin{equation}\label{eq:mf_int}
    \bar{\mathbf{u}}_{r,t}(\mathbf{x}_t\mid\mathbf{x}_1)
    =
    \frac{1}{t-r}
    \int_r^t
    \mathbf{u}_\tau(\mathbf{x}_\tau\mid\mathbf{x}_1)\,\mathrm{d}\tau .
\end{equation}
This enables single-step transport,
\begin{equation}
    \mathbf{x}_r
    =
    \mathbf{x}_t-(t-r)\bar{\mathbf{u}}_{r,t}(\mathbf{x}_t\mid\mathbf{x}_1).
\end{equation}
Recent work improves MF stability near the flow-matching limit~\cite{geng2025improved, zhang2025alphaflow}. Pixel-space variants remove the VAE bottleneck, building on latent-free image-generation architectures~\cite{li2025back}. In particular, pixel MeanFlow (p-MF) directly predicts in image space and enables one-step generation~\cite{lu2026onesteplatentfreeimagegeneration}.

\paragraph{Fast diffusion samplers.}
Even when training uses DDPM-style objectives, inference can be accelerated by solving the PF-ODE with better numerical integrators. DDIM is a first-order deterministic sampler~\cite{song2020denoising}. DPM-Solver and DPM++ formulate the PF-ODE in log-SNR coordinates and use exponential integrators~\cite{lu2022dpmsolverfastodesolver, lu2025dpm}. DEIS similarly uses polynomial interpolation of previous model predictions to construct high-order exponential integrator updates~\cite{zhang2022fast}. These samplers provide strong few-step baselines for comparing against one-step p-MF.

\section{Methodology}\label{sec:methodology}

\subsection{Problem Setup}

We study redshift-conditioned galaxy image generation on GalaxiesML-64~\cite{do2024galaxiesmldatasetgalaxyimages}. Each training example consists of an image $\mathbf{x}$ and a continuous redshift condition $c$. Our goal is to learn a conditional generator that samples realistic galaxy images $\hat{\mathbf{x}}\sim p(\mathbf{x}\mid c)$ while reducing inference cost relative to many-step DDPM sampling.

Our diffusion baseline follows \citet{lizarraga2025understandinggalaxymorphologyevolution}, which uses DDPMs for redshift-conditioned galaxy morphology generation. We compare this baseline against accelerated diffusion samplers and a one-step p-MF model.

\subsection{Redshift-conditioned pixel MeanFlow}
We adapt p-MF~\cite{lu2026onesteplatentfreeimagegeneration} to continuous redshift conditioning. p-MF builds on MF models, which directly model the average velocity over a time interval as a self-supervised signal in pixel-space. Given a clean image $\mathbf{x}$, Gaussian noise $\boldsymbol{\epsilon}\sim\mathcal{N}(0,I)$, and time $t$, we use the flow-matching (FM) interpolation path
\begin{equation}
    \mathbf{z}_t=(1-t)\mathbf{x}+t\boldsymbol{\epsilon}.
\end{equation}
The neural network backbone is an attention-augmented NCSN-style U-Net~\cite{song2019generative}, implemented in \texttt{jax}~\cite{jax2018github} and \texttt{flax}~\cite{flax2020github}. It predicts a clean-image parameterization
$D_\theta(\mathbf{z}_t,r,t,w,c)$, where $r$ is the reference time, $w$ is the classifier-free guidance weight, and $c$ denotes redshift. The induced average FM velocity is
\begin{equation}
    \mathbf{v}_\theta(\mathbf{z}_t,r,t,w,c)
    =
    \frac{1}{t}\left[\mathbf{z}_t-D_\theta(\mathbf{z}_t,r,t,w,c)\right].
\end{equation}
To incorporate classifier-free guidance during training, we compute conditional and unconditional velocities and define
\begin{equation}
    \mathbf{v}_{\mathrm{guide}}
    =
    (\boldsymbol{\epsilon}-\mathbf{x})
    +
    \left(1-\frac{1}{w}\right)
    \left(\mathbf{v}_{\mathrm{cond}}-\mathbf{v}_{\mathrm{uncond}}\right),
\end{equation}
where the first term is the optimal-transport FM velocity and the second term applies training-time guidance. Following the improved MeanFlow formulation~\cite{geng2025improved}, we use the MeanFlow identity, obtained by differentiating Eq.~\eqref{eq:mf_int}, to construct the training target for the induced instantaneous velocity $\mathbf{V}_\theta$:
\begin{equation}
    \mathbf{V}_\theta
    =
    \mathbf{v}_{\mathrm{cond}}
    +
    (t-r)\operatorname{sg}\!\left(
    \partial_t\mathbf{v}_{\mathrm{cond}}
    +
    \mathbf{v}_{\mathrm{cond}}^{\top}\nabla_z\mathbf{v}_{\mathrm{cond}}
    \right).
\end{equation}
We then optimize the FM-style velocity regression loss
\begin{equation}\label{eq:pmf_loss}
    \mathcal{L}_{\mathrm{p\text{-}MF}}
    =
    \mathbb{E}
    \left[
    \left\|\mathbf{V}_\theta-\mathbf{v}_{\mathrm{guide}}\right\|_2^2
    \right].
\end{equation}
At inference time, generation requires a single neural function evaluation (NFE):
\begin{equation}\label{eq:pmf_inference}
    \hat{\mathbf{x}}
    =
    \boldsymbol{\epsilon}
    -
    \mathbf{v}_{\theta^*}(\boldsymbol{\epsilon},0,1,w,c),
    \qquad
    \boldsymbol{\epsilon}\sim\mathcal{N}(0,I).
\end{equation}
The training and inference procedures are summarized in Algorithms~\ref{alg:training} and~\ref{alg:inference}.

\begin{figure*}[h]
    \centering
    \includegraphics[width=1.\linewidth]{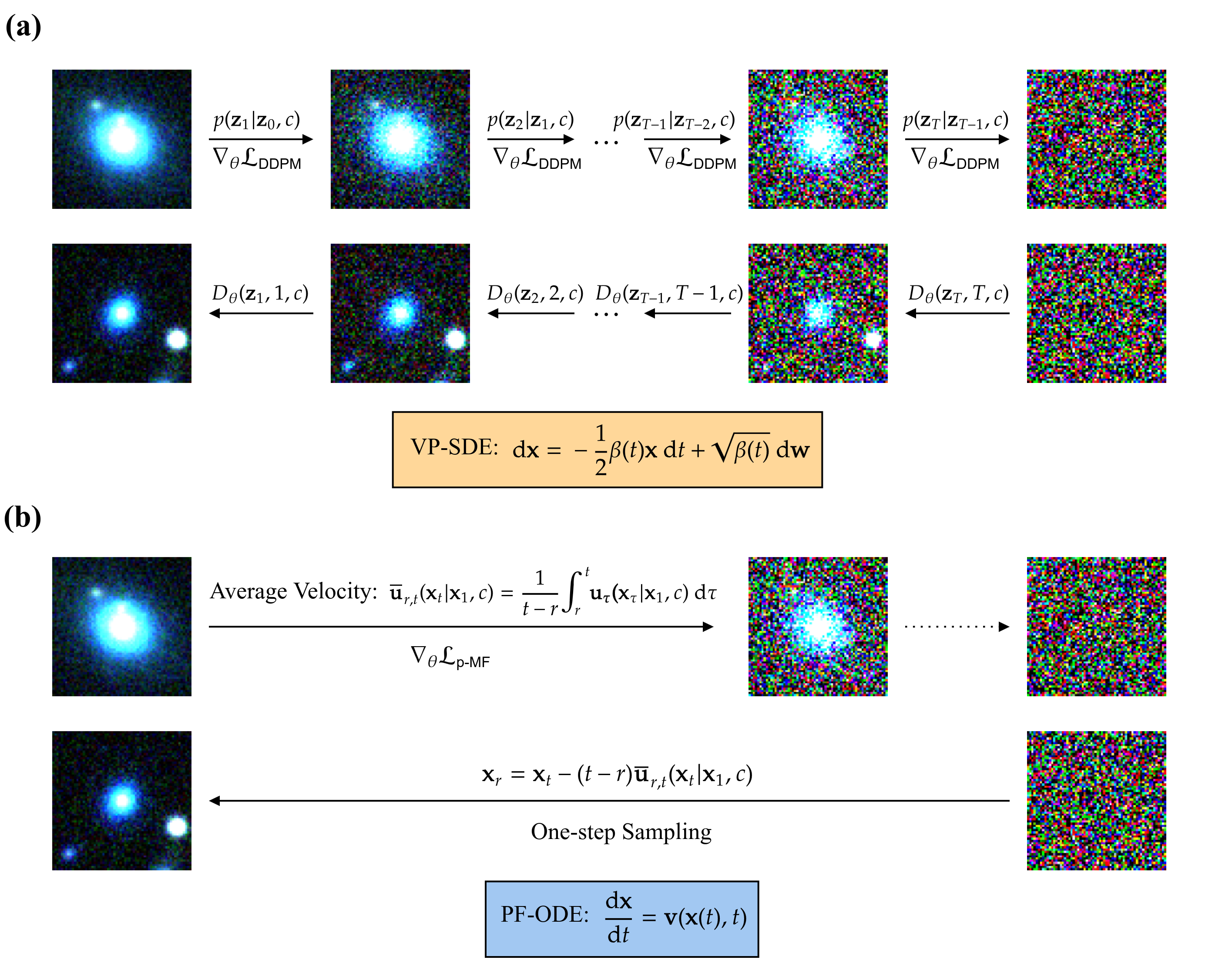}
    \caption{Comparison between the DDPM baseline of \citet{lizarraga2025understandinggalaxymorphologyevolution} and our p-MF model. DDPM uses multi-step reverse diffusion, whereas p-MF learns an average velocity field and generates in one step.}
    \label{fig:comparison}
    \vspace{-1em}
\end{figure*}

\subsection{Baselines and Evaluation}

We compare p-MF against a DDPM model sampled with four solvers: standard DDPM sampling, deterministic DDIM with $\eta=0$~\cite{song2020denoising}, DPM++2M~\cite{lu2025dpm}, and an AB2-style DEIS sampler~\cite{zhang2022fast}. This allows us to separate the effects of model class from the effects of sampler efficiency.

All models are evaluated following the protocol of \citet{lizarraga2025understandinggalaxymorphologyevolution}. We measure agreement between generated and observed galaxies using morphology-based statistics: ellipticity, semi-major axis, S\'ersic index, and isophotal area. We report both unconditional distributional agreement and redshift-conditioned agreement, enabling a direct comparison between many-step diffusion sampling, few-step diffusion solvers, and one-step p-MF generation. We also ablate key design choices from \citet{lizarraga2025understandinggalaxymorphologyevolution}, including redshift label smoothing and robust loss functions, in the p-MF setting.

\begin{algorithm}[h]
\caption{pixel-MeanFlow Training}
\label{alg:training}
\begin{algorithmic}[1]
\REQUIRE dataset $\mathcal{D}$ (empirical distribution $\hat p_{\mathcal D}$), denoiser $D_\theta$, iterations $K$, batch size $B$, learning rate $\eta$
\STATE Initialize $\theta$
\FOR{$k=1,\dots,K$}
    \STATE Sample $\{\mathbf{x}^{(i)}, c^{(i)}\}_{i=1}^B \sim \hat p_{\mathcal D}$, $\boldsymbol{\epsilon}^{(i)}\sim\mathcal N(0,I)$, $t^{(i)}$, $r^{(i)}$, $w^{(i)}$
    \STATE $\mathbf{z}_{t}^{(i)}\gets (1-t^{(i)})\mathbf{x}^{(i)}+t^{(i)}\boldsymbol{\epsilon}^{(i)}$
    \STATE \# \textit{Sample Velocities}
    \STATE $\mathbf{v}_{\theta, \, \text{uncond}}^{(i)} \gets \frac{1}{t^{(i)}}\left[\mathbf{z}_t^{(i)} - D_\theta(\mathbf{z}_t^{(i)},r^{(i)},t^{(i)}, w^{(i)}, \varnothing)\right]$
    \STATE $\mathbf{v}_{\theta, \, \text{cond}}^{(i)} \gets \frac{1}{t^{(i)}}\left[\mathbf{z}_t^{(i)} - D_\theta(\mathbf{z}_t^{(i)},r^{(i)},t^{(i)}, w^{(i)}, c^{(i)})\right]$
    \STATE $\mathbf{v}_{\theta, \, \text{guidance}}^{(i)} \gets \left(\boldsymbol{\epsilon}^{(i)} - \mathbf{x}^{(i)}\right) + \left(1 - \frac{1}{w^{(i)}} \right) (\mathbf{v}_{\theta, \, \text{cond}}^{(i)} - \mathbf{v}_{\theta, \, \text{uncond}}^{(i)})$
    \STATE \# \textit{MeanFlow Identity}
    \STATE $\mathbf{V}_\theta^{(i)}\gets \mathbf{v}_{\theta, \, \text{cond}}^{(i)}+(t^{(i)}-r^{(i)})\cdot\text{sg}\Big(\partial_t \mathbf{v}_{\theta, \, \text{cond}}^{(i)}+\mathbf{v}_{\theta, \, \text{cond}}^{(i)\,\top}\nabla_{z}\mathbf{v}_{\theta, \, \text{cond}}^{(i)}\Big)$
    \STATE $\mathcal{L}_{\text{p-MF}}\gets \frac{1}{B}\sum_{i=1}^B \left\|\mathbf{V}_\theta^{(i)}-\mathbf{v}_{\theta, \, \text{guidance}}^{(i)}\right\|_2^2$
    \STATE $\theta \gets \theta-\eta\nabla_\theta\big(\mathcal{L}_{\text{p-MF}}\big)$
\ENDFOR
\end{algorithmic}
\end{algorithm}

\begin{algorithm}[t]
\caption{pixel-MeanFlow Inference}
\label{alg:inference}
\begin{algorithmic}[1]
\REQUIRE trained models $D_{\theta^*}$, condition $\{c^{(i)}\}_{i=1}^B$, guidance strength $\{w^{(i)}\}_{i=1}^B$
\STATE Sample $\boldsymbol{\epsilon}^{(i)} \sim \mathcal{N}(0,I)$
\STATE $\hat{\mathbf{x}}^{(i)} \leftarrow \boldsymbol{\epsilon}^{(i)} - \mathbf{v}^{(i)}_{\theta^*}(\boldsymbol{\epsilon}^{(i)}, 0, 1, w^{(i)}, c^{(i)})$
\end{algorithmic}
\textbf{Return:} generated images $\{\hat{\mathbf{x}}^{(i)}\}_{i=1}^B$.
\end{algorithm}

\section{Experiments}
\paragraph{Baselines.} In this section, we evaluate p-MF and DDPM models using four samplers: standard DDPM with $1000$ sampling steps, DDIM with $\eta=0$ and $60$ sampling steps, DPM++2M with $30$ sampling steps, and DEIS-AB2 with $30$ sampling steps. The DDPM model with the standard DDPM sampler, implemented following \citet{lizarraga2025understandinggalaxymorphologyevolution}, serves as the baseline for comparison with our p-MF models. See Appendix \ref{sec:exp_settings} for details.

\paragraph{Metrics.} We assess generation quality using two types of metrics. First, we compute the Jensen--Shannon divergence (JSD) between the unconditional distributions of physical quantities measured from real and generated samples, including ellipticity, semi-major axis, S\'ersic index, and isophotal area. Second, we evaluate the normalized Bin Center Differences (BCD) between real and generated samples under redshift-conditioned generation. These physical quantities are computed using the implementation of \citet{lizarraga2025understandinggalaxymorphologyevolution}. All evaluations are performed on the official test split of the GalaxiesML-64 dataset~\cite{do2024galaxiesmldatasetgalaxyimages}. See Appendix \ref{sec:metrics} for details.

\paragraph{Results.}
The experimental results are summarized in Table~\ref{tab:result_table}. We report conditional binned distributional discrepancies (BCD) and unconditional Jensen--Shannon divergences (JSD) for four astrophysical morphology statistics: semi-major axis (SMA), S\'ersic index (SI), isophotal area (IA), and ellipticity (Ell.). Lower values indicate better agreement between the generated and observed galaxy distributions.

Among diffusion-based methods, the standard DDPM sampler achieves the best overall performance. This is expected, since DDPM uses $1000$ reverse steps and therefore most closely follows the original reverse diffusion process. It obtains the lowest unconditional JSD across all four quantities and substantially outperforms the accelerated samplers in conditional BCD, especially for SI, IA, and ellipticity. This indicates that, although the DDPM sampler is computationally expensive, it remains the strongest baseline in terms of distributional fidelity.

\begin{table}[h] 
\centering 
\small 
\caption{ Summary of the main model results. Uncertainty estimates obtained by bootstrapping are reported in parentheses. Distribution-level metrics are shown for four key morphological parameters. The one-step p-MF model preserves reasonable morphology statistics while requiring three orders of magnitude fewer function evaluations than DDPM.} 
\setlength{\tabcolsep}{0.3em} 
\begin{tabular}{lc|cccc|cccc} 
\toprule 
\multicolumn{2}{c|}{\textbf{Model Specs}} & \multicolumn{4}{c|}{\textbf{Conditional BCD}$\downarrow$} & \multicolumn{4}{c}{\textbf{Unconditional JSD}$\downarrow$} \\ \midrule \textbf{Model Name} & \textbf{NFE} & \textbf{SMA} & \textbf{SI} & \textbf{IA} & \textbf{Ell.} & \textbf{SMA} & \textbf{SI} & \textbf{IA} & \textbf{Ell.} \\ \midrule \rowcolor{gray!15} \multicolumn{10}{l}{\textit{Diffusion Models}}\\ DDPM & 1000 & 0.249\unc{40} & 1.37\unc{50} & 0.414\unc{54} & 0.460\unc{91} & 0.0036\unc{4} & 0.0003\unc{1} & 0.0045\unc{4} & 0.0049\unc{4} \\ DDIM ($\eta=0$) & 60 & 2.99\unc{23} & 11.3\unc{8} & 2.54\unc{14} & 6.83\unc{38} & 0.0305\unc{16} & 0.0062\unc{6} & 0.1832\unc{33} & 0.0779\unc{23} \\ DEIS-AB2 & 30 & 0.368\unc{75} & 5.69\unc{39} & 1.94\unc{11} & 1.06\unc{21} & 0.0142\unc{12} & 0.0011\unc{3} & 0.0776\unc{25} & 0.0311\unc{17} \\ DPM++2M & 30 & 0.328\unc{110} & 5.80\unc{45} & 2.53\unc{17} & 1.06\unc{29} & 0.0497\unc{19} & 0.0033\unc{5} & 0.0682\unc{22} & 0.0227\unc{14} \\ \rowcolor{gray!15} \multicolumn{10}{l}{\textit{p-MF Models}}\\ p-MF (25M, L$_2$) & 1 & 0.462\unc{87} & 9.03\unc{71} & 2.18\unc{6} & 3.33\unc{34} & 0.0947\unc{33} & 0.0069\unc{9} & 0.2126\unc{27} & 0.0311\unc{22} \\ \bottomrule 
\end{tabular} 
\label{tab:result_table} 
\end{table}

For efficient diffusion sampling, the second-order solvers consistently improve over the first-order DDIM sampler. DDIM with $\eta=0$ and $60$ NFEs exhibits a clear degradation relative to standard DDPM, particularly in the conditional metrics. In contrast, DEIS-AB2 and DPM++2M use only $30$ NFEs but achieve better results than DDIM on most quantities. For example, both second-order solvers substantially reduce the conditional BCD for SMA, SI, and ellipticity. DEIS-AB2 gives the best overall performance among the efficient solvers, achieving lower conditional BCD for SI and IA and lower unconditional JSD for SMA and SI, while DPM++2M performs slightly better on unconditional IA and ellipticity. These results suggest that higher-order exponential-integrator-based solvers are more effective than first-order DDIM in the low-NFE regime.

\begin{figure*}[h]
    \centering
    \includegraphics[width=\linewidth]{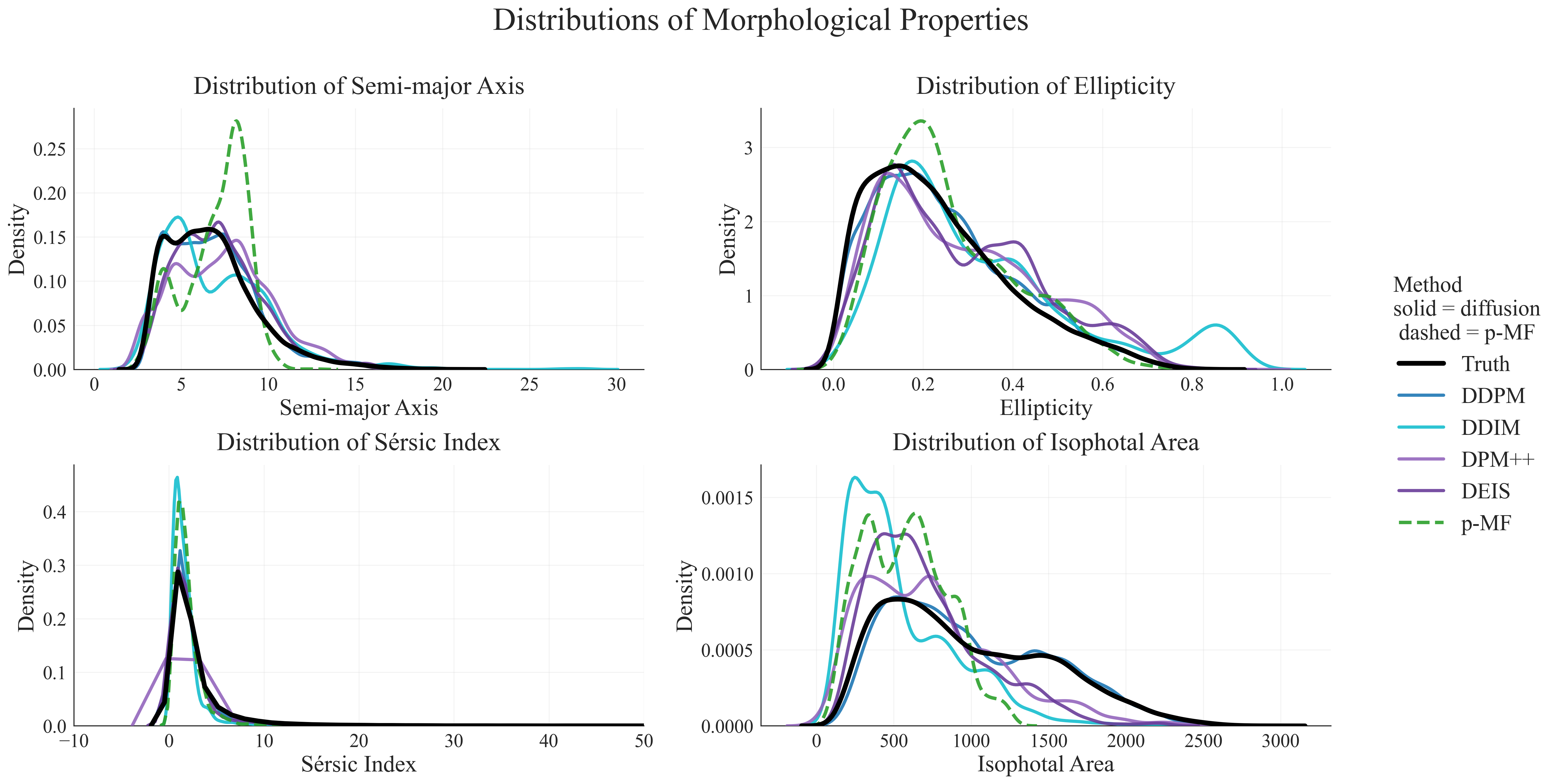}
    \caption{
    Unconditional morphological distributions for all models. The p-MF model is shown with green dashed lines.
    }
    \label{fig:uncon_dist}
    \vspace{-1em}
\end{figure*}

We also compare the diffusion baselines against a $25$M-parameter p-MF model, which requires only a single NFE at inference time. p-MF does not match the full $1000$-step DDPM sampler in generation quality. Nevertheless, despite using only one inference step, p-MF achieves metrics that are competitive with accelerated diffusion samplers on several quantities. In particular, its conditional IA is comparable to DEIS-AB2 and better than DDIM and DPM++2M, while its unconditional ellipticity JSD is comparable to DEIS-AB2. The main degradation appears in the S\'ersic index and unconditional SMA/IA distributions, suggesting that one-step p-MF captures coarse morphology reasonably well but still struggles with some fine-grained structural statistics. A qualitative inspection of the generated samples suggests that these metric degradations are associated with recognizable morphological failure modes. In particular, the one-step p-MF samples sometimes show oversmoothing and PSF-like blurring, which can wash out compact central cores and make disk components appear more diffuse. We also observe cases with noisier low-surface-brightness outskirts, imperfect preservation of ellipticity, and occasional colour artifacts, indicating that the model does not always retain fine-scale structural and photometric details. These effects are consistent with a limited one-step generator capturing the dominant modes of the galaxy distribution while under-representing rarer or sharper morphological features, suggesting mild mode collapse in the highest-detail regimes rather than a complete failure of global morphology modeling (See Appendix \ref{appendix:more_results} for the visual distinction).

The computational trade-off is shown in Table~\ref{tab:infer_table}. Standard DDPM requires $33.5$ seconds and $7020$ GFLOPs for inference, making it substantially more expensive than all accelerated alternatives. DDIM reduces inference time to $2.6$ seconds, while DEIS-AB2 and DPM++2M further reduce it to approximately $1.5$--$1.6$ seconds. The $25$M p-MF model achieves a comparable inference time of $1.5$ seconds with only one network evaluation and a lower total GFLOP count than the $30$-step diffusion solvers. The smaller $1$M p-MF model further reduces inference time to $0.9$ seconds and requires only $23.2$ GFLOPs. Overall, these results highlight a clear accuracy--efficiency trade-off: standard DDPM provides the best fidelity, second-order diffusion solvers offer strong performance at much lower cost, and p-MF provides a promising one-step alternative with substantial computational savings. The comparison of unconditional distributions for different morphological parameters is shown in Figure \ref{fig:uncon_dist}.

\begin{table}[h] 
\centering 
\caption{ Summary of inference speed (in seconds), GFLOPs, and model size for all models recorded on a single NVIDIA 4090 GPU. For diffusion models, the GFLOP count is computed by multiplying the single-step FLOP count by the number of solver steps. All samplers are compiled at the loop level using \texttt{jax.jit}. } 
\setlength{\tabcolsep}{0.5em} 
\begin{tabular}{l|cccc} 
\toprule \textbf{Model Name} & \textbf{NFE} & \textbf{Size} & \textbf{Inference Speed} & \textbf{GFLOPs} \\ \midrule \rowcolor{gray!15} \multicolumn{5}{l}{\textit{Diffusion Models}}\\ DDPM & 1000 & 21.4M & 33.5\unc{6} & 7020 \\ DDIM ($\eta=0$) & 60 & 21.4M & 2.6\unc{4} & 421 \\ DEIS-AB2 & 30 & 21.4M & 1.6\unc{3} & 197 \\ DPM++2M & 30 & 21.4M & 1.5\unc{4} & 198 \\ \rowcolor{gray!15} \multicolumn{5}{l}{\textit{p-MF Models}}\\ p-MF (1M) & 1 & 1.26M & 0.9\unc{2} & 23.2 \\ p-MF (25M) & 1 & 25.2M & 1.5\unc{1} & 154 \\ 
\bottomrule 
\end{tabular} 
\label{tab:infer_table} 
\end{table}

\paragraph{Ablation Studies.}
\citet{lizarraga2025understandinggalaxymorphologyevolution} proposed two practical modifications for astrophysical diffusion models: label smoothing on the redshift conditioning variable and replacing the standard DDPM $L_2$ objective with the Huber loss. We investigate whether these design choices remain effective in the context of one-step p-MF models. We perform ablations at two model scales, $1$M and $25$M parameters. In Table~\ref{tab:ablation_table}, NLS denotes the variant without label smoothing.

The ablation results are shown in Table~\ref{tab:ablation_table}. For the $1$M p-MF models, the effect of the Huber loss is mixed. Compared with the $L_2$ objective, Huber loss improves the conditional BCD for SMA and slightly improves the unconditional JSD for SI and ellipticity. However, it substantially worsens the IA metrics and also degrades the unconditional SMA distribution. The $L_2$ objective therefore provides the most balanced performance at the $1$M scale, achieving the best conditional IA and ellipticity, as well as the best unconditional SMA and IA.

Removing label smoothing also has non-uniform effects. In the $1$M setting, comparing the Huber and Huber--NLS variants shows that removing label smoothing improves conditional SI, conditional IA, and unconditional ellipticity, but significantly worsens conditional SMA. This suggests that label smoothing may help stabilize the conditioning signal for certain morphology statistics, while potentially oversmoothing others.

At the larger $25$M scale, the trade-off becomes clearer. The $25$M p-MF model trained with the $L_2$ objective achieves the best conditional BCD across all four quantities, indicating that label smoothing is beneficial for redshift-conditioned generation. In contrast, the $25$M model without label smoothing obtains better unconditional JSD across all four quantities. This indicates that removing label smoothing improves the marginal realism of generated samples, but at the cost of weaker conditional alignment.

Overall, these ablations suggest that design choices originally proposed for DDPMs do not transfer uniformly to p-MF models. The Huber loss improves some individual morphology statistics but does not consistently outperform the simpler $L_2$ objective. Label smoothing appears more useful for conditional generation, particularly at larger model scale, whereas removing it can improve unconditional distributional matching. Therefore, in our main comparison, we use the $25$M p-MF model trained with the $L_2$ objective as the default setting, since it provides the strongest conditional performance and the best overall trade-off for redshift-conditioned galaxy generation.

\begin{table}[h]
\centering
\small
\caption{
Ablation study of p-MF models with different loss functions and redshift-conditioning strategies. NLS denotes training without label smoothing. Uncertainty estimates obtained by bootstrapping are reported in parentheses. 
}
\setlength{\tabcolsep}{0.3em}
\begin{tabular}{lc|cccc|cccc}
\toprule
\multicolumn{2}{c|}{\textbf{Model Specs}}  
& \multicolumn{4}{c|}{\textbf{Conditional BCD}$\downarrow$} 
& \multicolumn{4}{c}{\textbf{Unconditional JSD}$\downarrow$} \\
\midrule
\textbf{Model Name} & \textbf{NFE}  
& \textbf{SMA}
& \textbf{SI}
& \textbf{IA}
& \textbf{Ell.}
& \textbf{SMA}
& \textbf{SI}
& \textbf{IA}
& \textbf{Ell.}
 \\
\midrule

\rowcolor{gray!15} 
\multicolumn{10}{l}{\textit{1M p-MF Models}}\\

p-MF \scriptsize{(1M, Huber, NLS)}
& 1
& 1.47\unc{4}
& \textbf{10.1}\unc{6}
& 2.17\unc{4}
& 3.78\unc{11} 	
& 0.0749\unc{15}
& 0.0050\unc{4}
& 0.2428\unc{26}
& \textbf{0.0313}\unc{11}
 \\

p-MF \footnotesize{(1M, L$_2$)}
& 1
& 0.864\unc{45}
& 11.4\unc{6}
& \textbf{1.98}\unc{5}
& \textbf{3.55}\unc{14} 	
& \textbf{0.0282}\unc{14}
& 0.0041\unc{5}
& \textbf{0.1520}\unc{28}
& 0.0486\unc{19}
 \\

p-MF \footnotesize{(1M, Huber)}
& 1
& \textbf{0.597}\unc{24}
& 10.4\unc{5}
& 3.16\unc{4}
& 3.75\unc{12}
& 0.0806\unc{16}
& \textbf{0.0039}\unc{4}
& 0.3140\unc{28}
& 0.0345\unc{12}
\\
\rowcolor{gray!15} 
\multicolumn{10}{l}{\textit{25M p-MF Models}}\\

p-MF \footnotesize{(25M, L$_2$)}                 
& 1 
& \textbf{0.462}\unc{87}
& \textbf{9.03}\unc{71}
& \textbf{2.18}\unc{6}
& \textbf{3.33}\unc{34} 	
& 0.0947\unc{33} 	
& 0.0069\unc{9}
& 0.2126\unc{27}
& 0.0311\unc{22} 	
 \\

 p-MF \scriptsize{(25M, L$_{2}$, NLS)}              
& 1 
& 0.780\unc{49}
& 9.99\unc{67}
& 2.41\unc{6}
& 3.55\unc{18}
& \textbf{0.0748}\unc{25}
& \textbf{0.0045}\unc{6}
& \textbf{0.2022}\unc{28}
& \textbf{0.0293}\unc{16}
 \\
\bottomrule
\end{tabular}
\label{tab:ablation_table}
\end{table}

\section{Conclusion}
\paragraph{Conclusion.}
In this work, we studied efficient redshift-conditioned generative modeling for galaxy morphology using diffusion models and pixel-MeanFlow. We reviewed the theoretical connections between score-based diffusion, Flow Matching, one-step generative models, and modern diffusion samplers, and evaluated these methods on the GalaxiesML-64 dataset.

Our results show a clear accuracy--efficiency trade-off. The $1000$-step DDPM sampler achieves the best overall morphology fidelity, but at high computational cost. Efficient samplers such as DEIS-AB2 and DPM++2M substantially reduce inference time while outperforming first-order DDIM in the low-NFE regime. The one-step p-MF model does not yet match full DDPM quality, but achieves competitive performance on several morphology statistics with only a single neural function evaluation. These results suggest that one-step and few-step generative models are promising tools for fast astrophysical image generation.

\paragraph{Limitations and Future Work.}
This study has several limitations. First, our experiments are restricted to low-resolution GalaxiesML-64 images, whereas real survey data include higher resolution, heterogeneous noise, point-spread-function effects, and selection biases. Second, our evaluation uses only a small set of morphology statistics, which may not fully capture image realism or physical consistency. Future work should include additional diagnostics, multi-band information, downstream scientific tasks, and expert visual assessment.

The current p-MF model also remains weaker than many-step DDPM sampling, especially for fine-grained structural statistics such as the S\'ersic index. Improving one-step generation may require larger models, longer training, improved MeanFlow objectives, or a small number of refinement steps. Finally, conditioning only on redshift limits the physical control of the model. Extending the framework to include additional galaxy properties such as stellar mass, star-formation rate, and environment would enable more detailed studies of galaxy evolution.

\bibliographystyle{unsrtnat}
\bibliography{ref}
\clearpage

\appendix

\section{Experimental Setting}\label{sec:exp_settings}
\paragraph{Optimization Setting.} 
The training setup for the diffusion model follows \citet{lizarraga2025understandinggalaxymorphologyevolution}. We use the Adam optimizer~\cite{kingma2014adam,goyal2017accurate} with a learning rate of $5\times 10^{-5}$. We also use the same exponential moving average (EMA) configuration as \citet{lizarraga2025understandinggalaxymorphologyevolution}. The diffusion model is trained on a single NVIDIA A100 80GB GPU for approximately $12$ hours. For the p-MF models, we train the 1M-parameter version for $170$ epochs on a single A100 GPU. For cost-matching purposes, we train the 25M-parameter version for $45$ epochs on $8$ NVIDIA GH200 GPUs for $8$ hours. All other optimization settings are kept the same as those used for the diffusion models. The hyperparameters are summarized in Table \ref{tab:optim_setting}.

\begin{table}[!h]
\centering
\caption{Optimization setting, same as \citet{lizarraga2025understandinggalaxymorphologyevolution}.}
\setlength{\tabcolsep}{0.3em}
\begin{tabular}{lcccc}
\toprule
\textbf{Name} & Learning Rate & EMA weight & Gradient Clipping & 
Learning Rate Schedule\\
\midrule
  \textbf{Values} & $5 \times 10^{-5}$ & 0.995 & No & constant  \\
\bottomrule
\end{tabular}
\label{tab:optim_setting}
\end{table}

\paragraph{Diffusion \& MeanFlow Model Setting.}The diffusion model setup follows \citet{lizarraga2025understandinggalaxymorphologyevolution}, which is based on \citet{ho2020denoising}. The configuration is summarized in Table~\ref{tab:diffusion_setting}.
\begin{table}[!h]
\centering
\caption{Diffusion model setting, same as \citet{lizarraga2025understandinggalaxymorphologyevolution}.}
\setlength{\tabcolsep}{0.3em}
\begin{tabular}{lccccc}
\toprule
\textbf{Name} & $\beta_\text{start}$ & $\beta_\text{end}$ & Diffusion Steps & 
Loss & Classifier Free Guidance \\
\midrule
  \textbf{Values} & 0.0001 & 0.02 & 1000 & Huber & No \\
\bottomrule
\end{tabular}
\label{tab:diffusion_setting}
\end{table}
The p-MF model settings follow the ImageNet-$256$ experimental configuration of \citet{lu2026onesteplatentfreeimagegeneration}. The configuration is summarized in Table~\ref{tab:pmf_setting}.
\begin{table}[!h]
\centering
\caption{pixel-MeanFlow model setting, same as \citet{lu2026onesteplatentfreeimagegeneration}.}
\setlength{\tabcolsep}{0.4em}
\begin{tabular}{lcccc}
\toprule
\textbf{Name} & $t, r$ Sampler & $w_\text{max}$ & $t\neq r$ Percentage  & Label Dropout Rate \\
\midrule
  \textbf{Values} & logit-normal(0.8, 0.8) & 7 & 50\%  & 0.1 \\
\bottomrule
\end{tabular}
\label{tab:pmf_setting}
\end{table}

\paragraph{Model Architecture.}
The diffusion model backbone is a \texttt{jax}-reimplementation of the UNet used by \citet{lizarraga2025understandinggalaxymorphologyevolution}. It consists of three layers in each downsampling, bottleneck, and upsampling block, and does not include attention or residual blocks. The p-MF backbone is adapted from the NCSN~\cite{song2019generative}/NCSN++~\cite{song2020improved} style UNet. It includes residual blocks at each resolution level and incorporates a self-attention block near the bottleneck.

\clearpage

\section{Metrics}\label{sec:metrics}
\subsection{Redshift-binned Morphology Difference}
\label{sec:metrics_bcd}

To evaluate whether generated samples reproduce the redshift dependence of key morphological summary
statistics, we compute a \textbf{Bin Center Difference} (BCD) between generated and reference measurements.
The metric is evaluated independently for each morphology statistic
\[
q \in
\{\texttt{semi\_major\_axis},\ \texttt{ellipticity},\ \texttt{isophotal\_area},\ \texttt{sersic\_index}\}.
\]
Let
\[
\{(r_i, q_i, \hat{q}_i)\}_{i=1}^{N}
\]
denote the redshift, reference morphology value, and generated morphology value for the $i$th matched sample,
respectively.

\paragraph{Redshift binning.}
We partition the redshift interval $[0,4]$ into eight equally spaced bins,
\begin{equation}
0 = e_0 < e_1 < \cdots < e_8 = 4,
\end{equation}
with bin edges given by
\begin{equation}
e_k = \frac{k}{2}, \qquad k=0,\dots,8.
\end{equation}
For bin $k$, the corresponding sample set is
\begin{equation}
\mathcal{I}_k
=
\{i : e_k \le r_i < e_{k+1}\},
\end{equation}
with the final bin closed on the right, so that samples with $r_i=e_8$ are included.

\paragraph{Validity filtering.}
Before computing the metric, we discard samples with non-finite redshift, reference value, or generated value.
We also discard samples for which either the reference or generated morphology value is negative. Thus the valid
index set is
\begin{equation}
\mathcal{V}
=
\{i :
r_i, q_i, \hat{q}_i \ \text{are finite},\ q_i \ge 0,\ \hat{q}_i \ge 0
\}.
\end{equation}
All bin averages and normalization factors are computed only over samples in $\mathcal{V}$.

\paragraph{Binned morphology means.}
For each redshift bin, we compute the reference and generated bin centers as the empirical means
\begin{equation}
\mu_k^{\mathrm{ref}}
=
\frac{1}{|\mathcal{I}_k \cap \mathcal{V}|}
\sum_{i \in \mathcal{I}_k \cap \mathcal{V}} q_i,
\end{equation}
and
\begin{equation}
\mu_k^{\mathrm{gen}}
=
\frac{1}{|\mathcal{I}_k \cap \mathcal{V}|}
\sum_{i \in \mathcal{I}_k \cap \mathcal{V}} \hat{q}_i.
\end{equation}
Bins containing no valid samples are ignored when computing the final discrepancy.

\paragraph{Summed Bin Center Difference.}
The Bin Center Difference is defined as the sum of absolute differences between generated and reference
bin-center means:
\begin{equation}
\mathcal{E}_{\mathrm{BCD}}(q)
=
\sum_{k \in \mathcal{K}}
\left|
\mu_k^{\mathrm{gen}} - \mu_k^{\mathrm{ref}}
\right|,
\end{equation}
where $\mathcal{K}$ denotes the set of redshift bins for which both bin means are finite. Lower values indicate
better agreement between the generated and reference redshift-conditioned morphology trends.

\paragraph{Normalized Bin Center Difference.}
To make the metric comparable across morphology statistics with different physical scales, we report a
normalized Bin Center Difference,
\begin{equation}
\mathcal{E}_{\mathrm{nBCD}}(q)
=
\frac{
\mathcal{E}_{\mathrm{BCD}}(q)
}{
\bar{q}_{\mathrm{ref}}
},
\end{equation}
where
\begin{equation}
\bar{q}_{\mathrm{ref}}
=
\frac{1}{|\mathcal{V}|}
\sum_{i \in \mathcal{V}} q_i
\end{equation}
is the global mean of the reference morphology statistic after validity filtering. If
$\bar{q}_{\mathrm{ref}}$ is zero or non-finite, the normalized metric is undefined.

\paragraph{Bootstrap uncertainty.}
We estimate uncertainty in the normalized BCD using nonparametric bootstrapping. Given $N_{\mathrm{boot}}$
bootstrap replicates, each replicate is formed by sampling $|\mathcal{V}|$ matched triples
$(r_i,q_i,\hat{q}_i)$ with replacement from the valid sample set. For bootstrap replicate $s$, we compute
\begin{equation}
\mathcal{E}_{\mathrm{nBCD}}^{(s)}(q),
\qquad s=1,\dots,N_{\mathrm{boot}}.
\end{equation}
In our implementation, we use $N_{\mathrm{boot}}=1000$ bootstrap samples with a fixed random seed for
reproducibility. We report the observed normalized BCD together with the bootstrap mean and standard deviation.

\subsection{Jensen-Shannon Divergence}
\label{sec:metrics_jsd}

To compare one-dimensional distributions of generated and reference summary statistics, we compute the
\textbf{Jensen-Shannon Divergence} (JSD). Unlike the Kullback-Leibler divergence, the Jensen-Shannon
divergence is symmetric and remains well defined when comparing two normalized discrete probability
distributions with shared support.

Let
\[
p, q \in \mathbb{R}^{D}
\]
denote two nonnegative vectors, such as histograms or discretized empirical distributions of a reference and
generated statistic. Before computing the divergence, both vectors are normalized to sum to one:
\begin{equation}
\tilde{p}_i
=
\frac{p_i+\varepsilon}{\sum_{j=1}^{D}(p_j+\varepsilon)},
\qquad
\tilde{q}_i
=
\frac{q_i+\varepsilon}{\sum_{j=1}^{D}(q_j+\varepsilon)},
\end{equation}
where $\varepsilon \ge 0$ is an optional small constant added before normalization.

\paragraph{Mixture distribution.}
The midpoint distribution is defined as
\begin{equation}
m_i
=
\frac{1}{2}
\left(
\tilde{p}_i + \tilde{q}_i
\right).
\end{equation}

\paragraph{Kullback-Leibler divergence.}
For two normalized discrete distributions $a$ and $b$, the Kullback-Leibler divergence is
\begin{equation}
D_{\mathrm{KL}}(a \,\|\, b)
=
\sum_{i:a_i>0}
a_i
\log_b
\left(
\frac{a_i}{b_i}
\right),
\end{equation}
where the sum is taken only over entries with $a_i>0$. In our implementation, the logarithm base is set to
$b=2$, so the divergence is measured in bits.

\paragraph{Jensen-Shannon divergence.}
The Jensen-Shannon divergence is then computed as
\begin{equation}
D_{\mathrm{JS}}(\tilde{p},\tilde{q})
=
\frac{1}{2}
D_{\mathrm{KL}}(\tilde{p}\,\|\,m)
+
\frac{1}{2}
D_{\mathrm{KL}}(\tilde{q}\,\|\,m).
\end{equation}
Lower values indicate closer agreement between the generated and reference distributions. A value of zero
indicates identical normalized distributions.

\paragraph{Validity conditions.}
Both input vectors must be one-dimensional, have the same length, contain only nonnegative entries, and have
strictly positive sums after optional $\varepsilon$ smoothing. If either vector has zero total mass, the
divergence is undefined.

\section{More Results}\label{appendix:more_results}

\begin{figure*}[h]
    \centering
    \includegraphics[width=1.\linewidth]{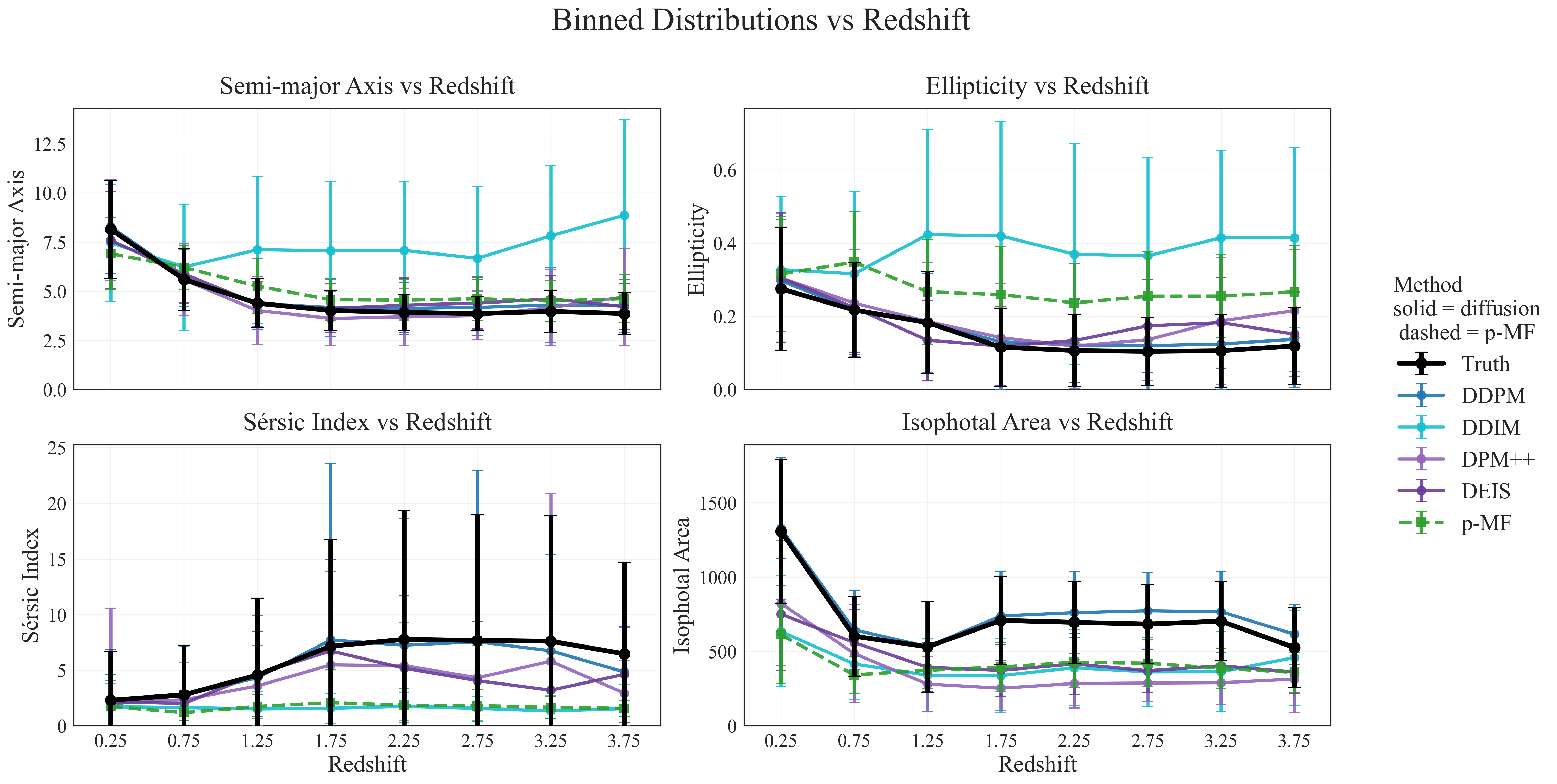}
    \caption{
    The redshift-conditional binned morphological distributions from all models, including p-MF model (in green dashed lines.)
    }
    \label{fig:fig4}
    \vspace{-1em}
\end{figure*}

\begin{figure*}[h]
    \centering
    \includegraphics[width=0.7\linewidth]{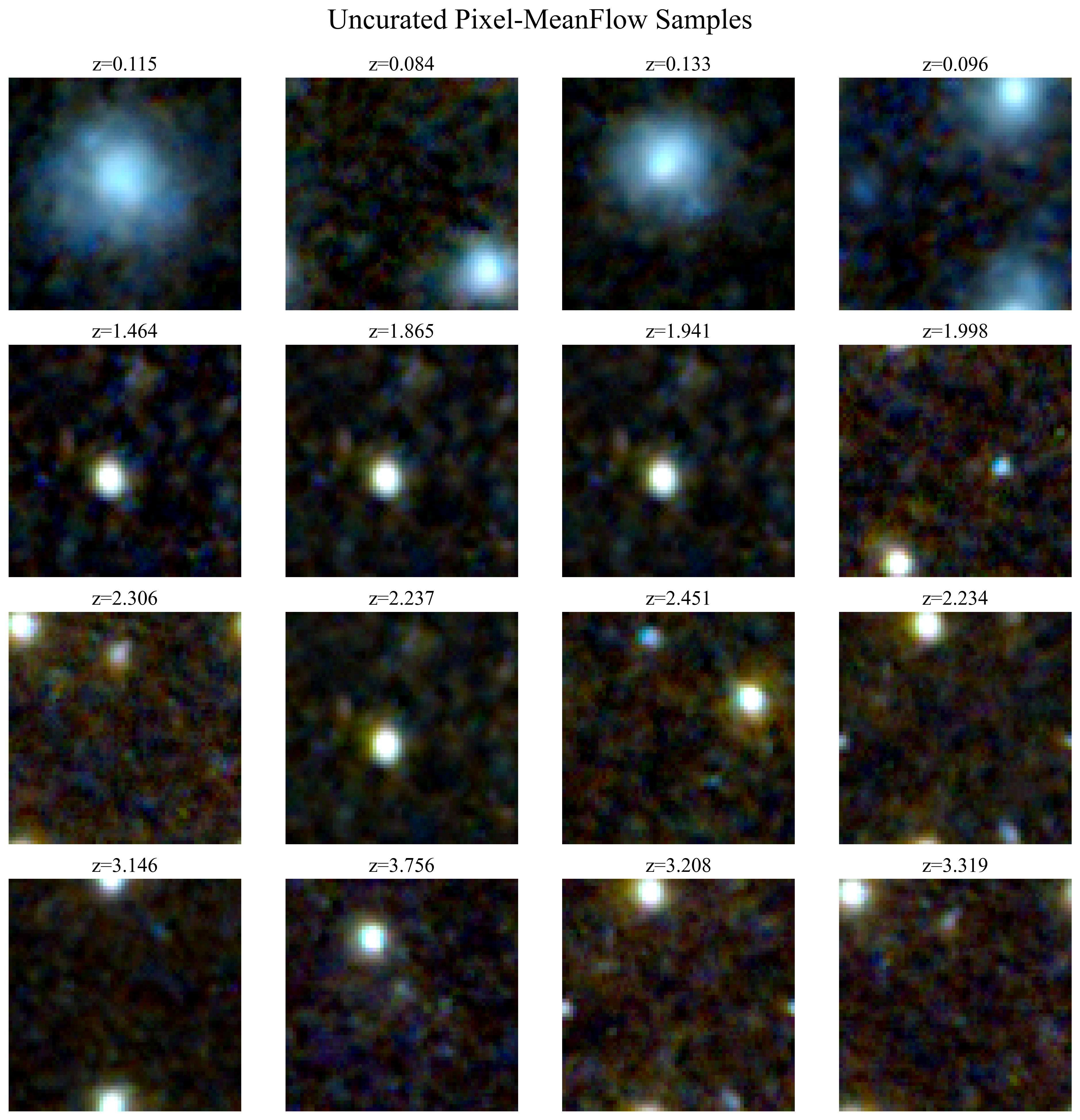}
    \caption{
    Uncurated samples from p-MF model in different redshift ranges. 
    }
    \label{fig:pmf_sample}
    \vspace{-1em}
\end{figure*}

\begin{figure*}[h]
    \centering
    \includegraphics[width=0.7\linewidth]{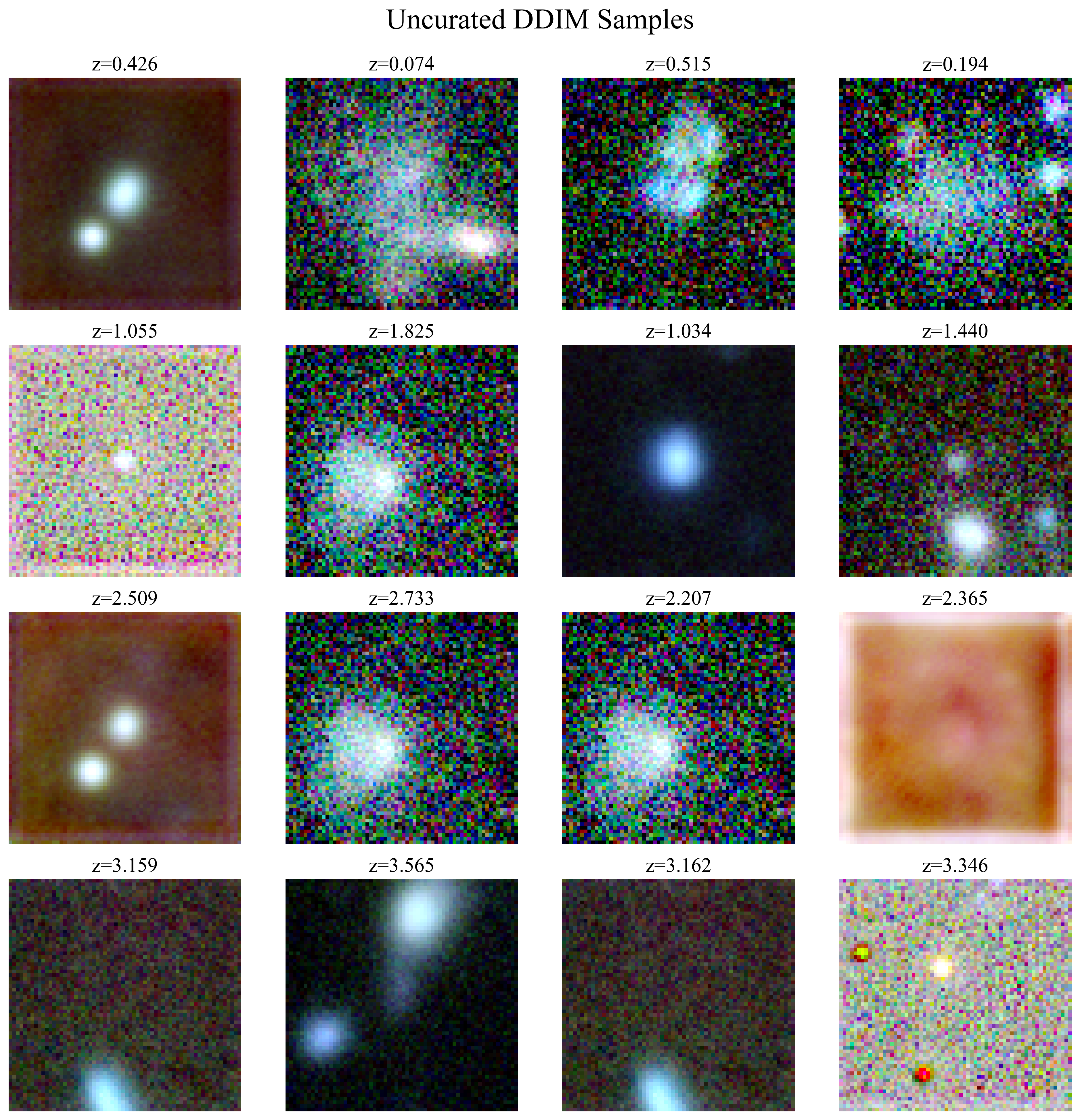}
    \caption{
    Uncurated samples from DDIM sampler in different redshift ranges. 
    }
    \label{fig:ddim_sample}
    \vspace{-1em}
\end{figure*}

\begin{figure*}[h]
    \centering
    \includegraphics[width=0.7\linewidth]{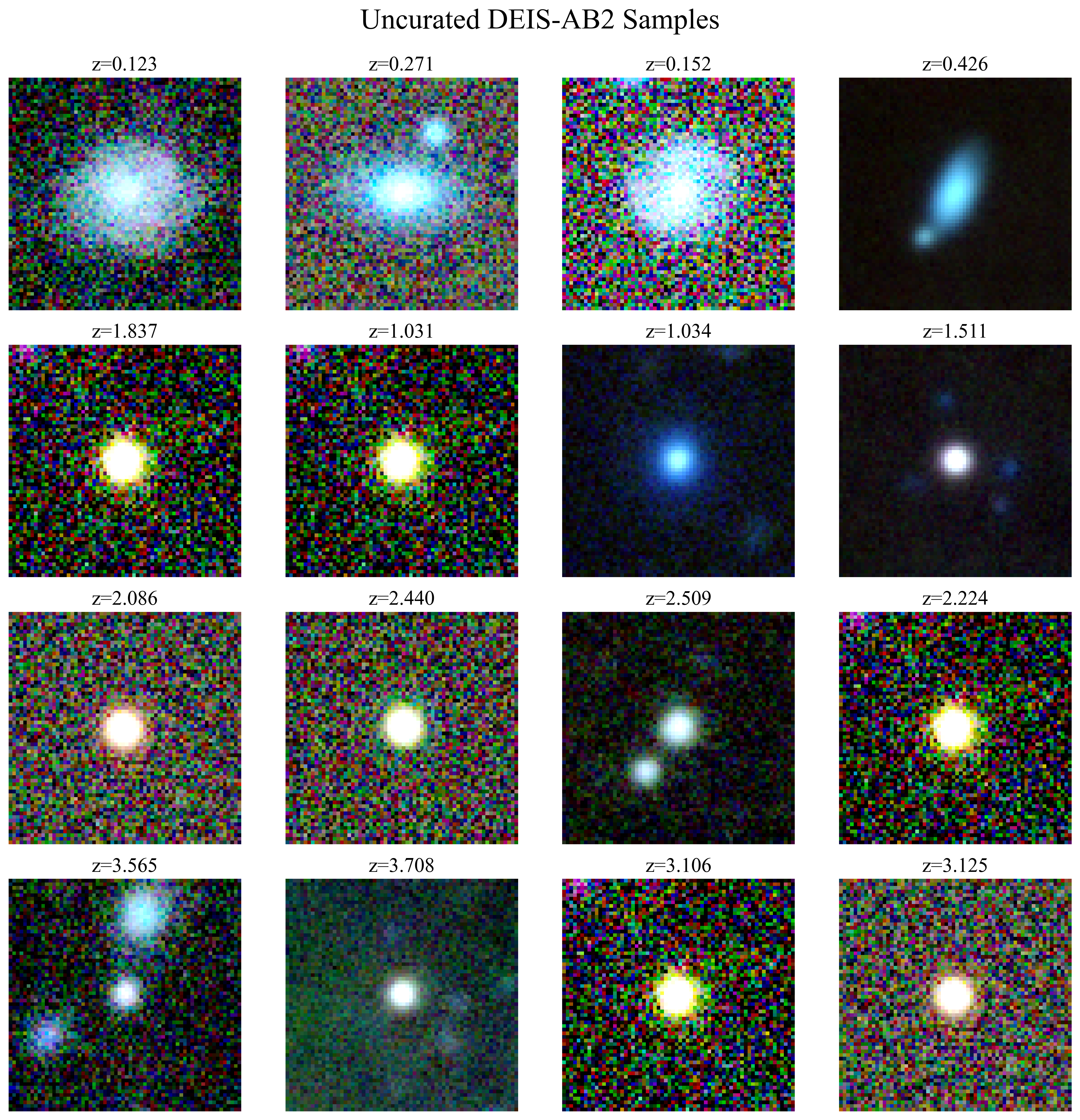}
    \caption{
    Uncurated samples from DEIS-AB2 sampler in different redshift ranges. 
    }
    \label{fig:deis_sample}
    \vspace{-1em}
\end{figure*}

\begin{figure*}[h]
    \centering
    \includegraphics[width=0.7\linewidth]{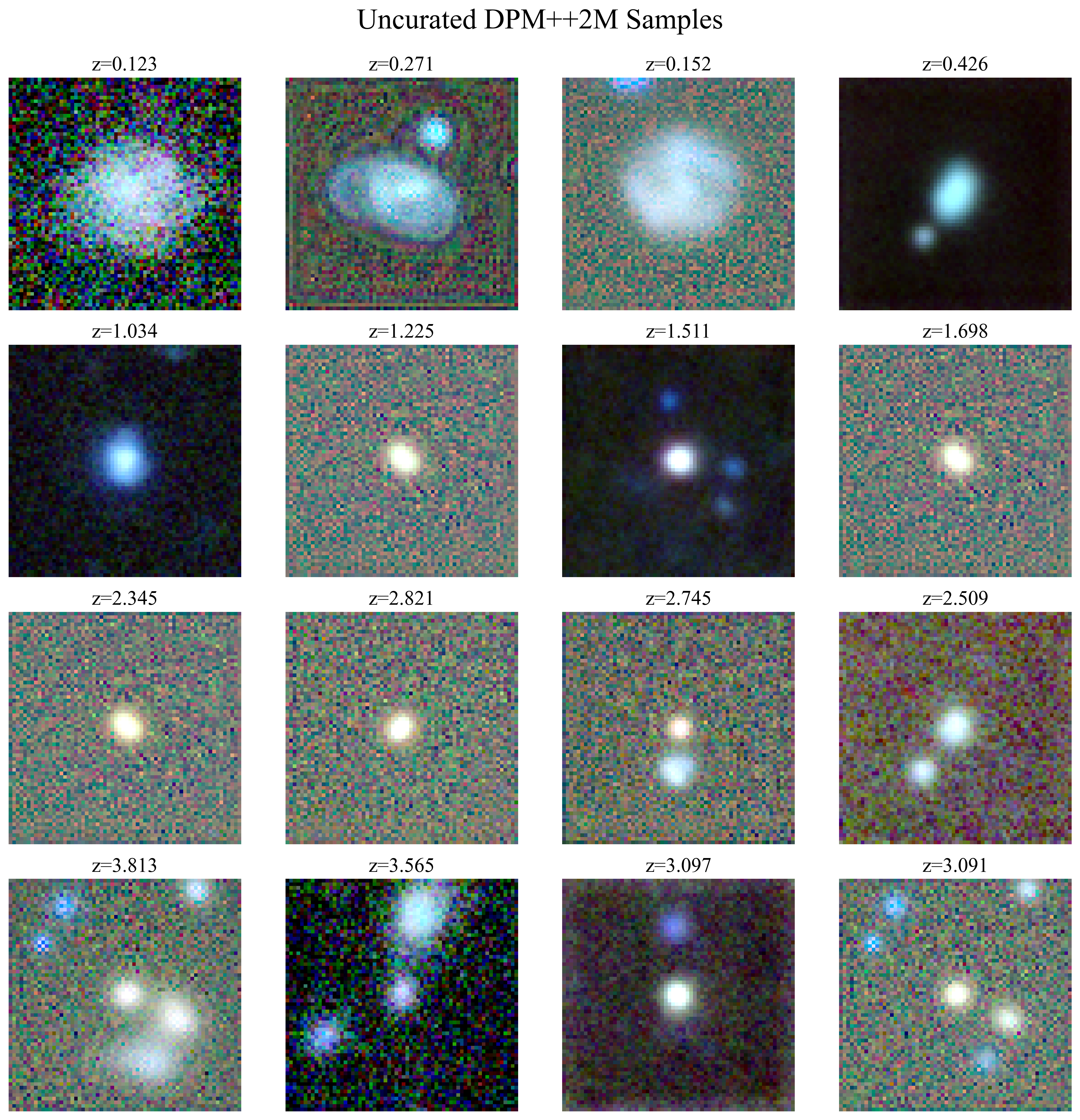}
    \caption{
    Uncurated samples from DPM++2M sampler in different redshift ranges. 
    }
    \label{fig:dpmpp_sample}
    \vspace{-1em}
\end{figure*}

\begin{figure*}[h]
\centering
\includegraphics[width=0.7\linewidth]{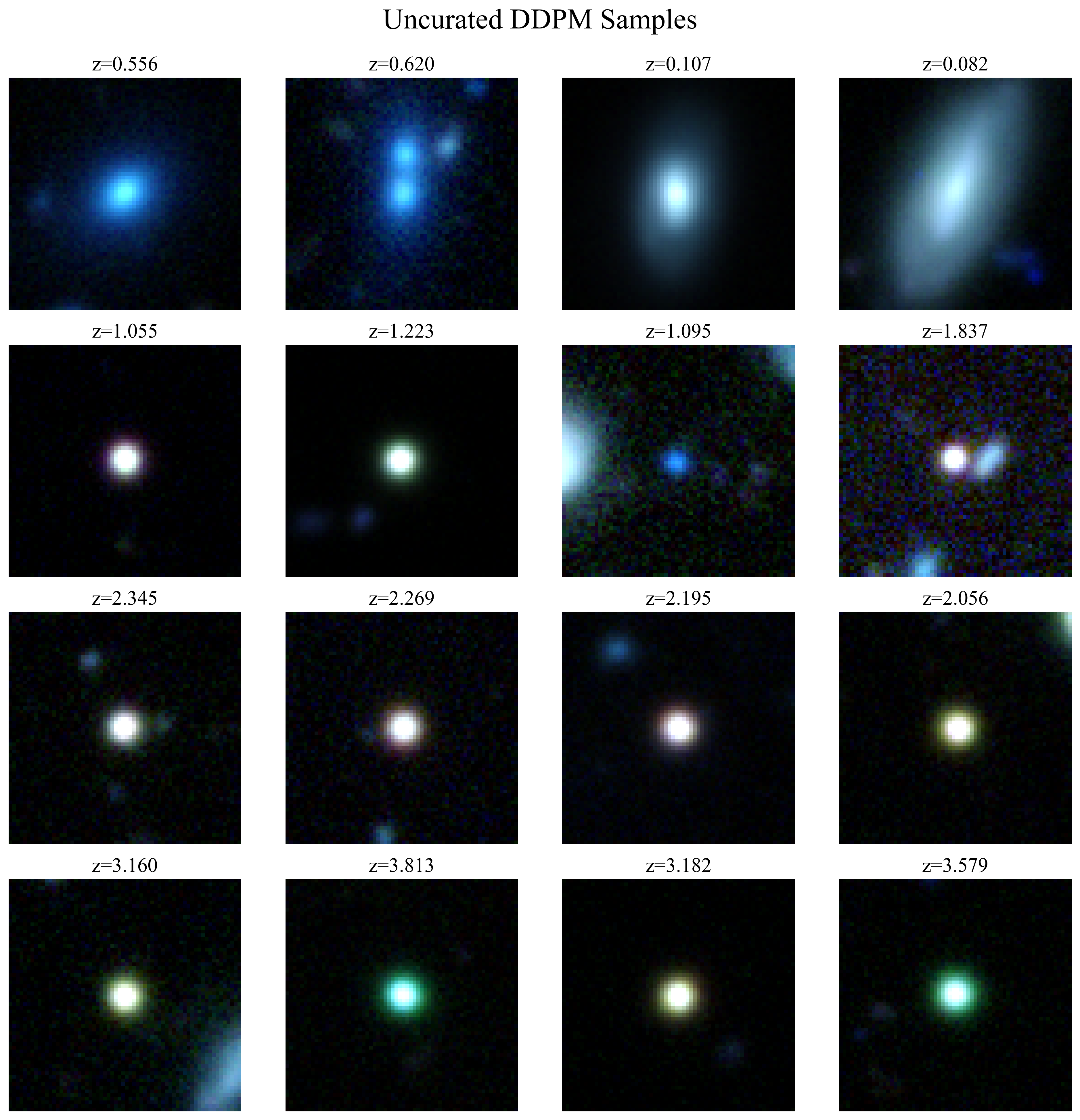}
\caption{
Uncurated samples from DDPM sampler in different redshift ranges. 
}
\label{fig:ddpm_sample}
\vspace{-1em}
\end{figure*}

\end{document}